\documentclass[%
 reprint,
 superscriptaddress,
 longbibliography,
 amsmath,amssymb,
 aps,
 prl,
]{revtex4-1}

\usepackage[colorlinks=true,citecolor=black]{hyperref}
\usepackage{graphicx}
\usepackage{dcolumn}
\usepackage{bm}
\usepackage[dvipsnames]{xcolor}
\usepackage[normalem]{ulem}
\usepackage{gensymb}

\begin{document}

\preprint{APS/123-QED}


\title{Angle resolved relaxation of spin currents by antiferromagnets in spin valves}

\author{D.~M.~Polishchuk}
\email{dpol@kth.se.}
\affiliation{Nanostructure Physics, Royal Institute of Technology, 10691 Stockholm, Sweden}%
\affiliation{Institute of Magnetism, NASU and MESU, 03142 Kyiv, Ukraine}

\author{A.~Kamra}
\affiliation{Center for Quantum Spintronics, Department of Physics, Norwegian University
of Science and Technology, NO-7491 Trondheim, Norway}

\author{T.~I.~Polek}
\affiliation{Institute of Magnetism,  NASU and MESU, 03142 Kyiv, Ukraine}
 
\author{A.~Brataas}%
\affiliation{Center for Quantum Spintronics, Department of Physics, Norwegian University
of Science and Technology, NO-7491 Trondheim, Norway}%

\author{V.~Korenivski}%
\affiliation{Nanostructure Physics, Royal Institute of Technology, 10691 Stockholm, Sweden}%

\date{\today}

\begin{abstract}

We observe and analyze tunable relaxation of a pure spin current by an antiferromagnet in spin-valves. This is achieved by carefully controlling the angle between a resonantly excited ferromagnetic layer pumping the spin current and the N\'eel vector of the antiferromagnetic layer. The effect is observed as an angle-dependent spin-pumping contribution to the ferromagnetic resonance linewidth. An interplay between spin-mixing conductance and, often disregarded, longitudinal spin conductance is found to underlie our observations, which is in agreement with a recent prediction for related ferromagnetic spin valves.

\end{abstract}

\maketitle

Spin polarization of the conduction electrons in metallic ferromagnets enables external control of the electrical properties of magnetic multilayers via the relative magnetization orientation of the ferromagnetic layers comprising the multilayer. The resulting angle-dependent transmission of a spin-polarized current is behind prominent effects such as giant~\cite{Baibich1988,Binasch1989} and tunneling ~\cite{Moodera1995} magnetoresistance (MR) as well as spin transfer torques~\cite{Slonczewski1996}, which paved the way for rapidly developing spin-electronics~\cite{Zutic2004,SpinCurrent2017}.

The related angle-dependent dissipation of spin-currents is behind the anisotropic~\cite{Park2011,Marti2014} and spin-Hall MR~\cite{Huang2012,Nakayama2013,Chen2016} observed in heavy-metal/antiferromagnet (HM/AF) bilayers~\cite{Han2014,Hoogeboom2017}. The key characteristic shared by these two effects is the change in the bilayer's resistance dependent on whether the polarization of the spin current in the HM is collinear or orthogonal to the axis of preferred spin alignment in the AF (its N\'eel vector $\mathbf{N}$). The demonstrated feasibility of controlling spin currents in antiferromagnetic nanostructures indicates a considerable potential of the emerging field of antiferromagnetic spintronics~\cite{Moriyama2015, Wadley2016, Jungwirth2016, Baltz2018}. However, an explicit, angle-resolved experimental study of the interaction between a spin current and the N\'eel vector of an AF, as well as its functional form and physical parameter space, that would underpin the existing strong theoretical effort~\cite{Nunez2006,Gomonay2010,Gomonay2018} is still pending.

Ferromagnetic resonance (FMR) driven spin pumping is a unique tool for analyzing spin relaxation in F/N/F$_\mathrm{st}$~\cite{Heinrich2003,Taniguchi2008,Ghosh2012,Marcham2013,Baker2016} and F/N/AF~\cite{Merodio2014,Frangou2016,Qiu2016,Moriyama2017} spin valves. The ferromagnetic layer F is the source as well as the probe of a pure spin-current pumped into the nonmagnetic spacer N and static ferromagnetic F$_\mathrm{st}$ or AF layers. Since there is negligible spin dissipation in the typically nm-thin spacer N, spin pumping probes the spin relaxation due to the static F$_\mathrm{st}$ or AF layers measured via the backflow spin-current contribution to the FMR linewidth of F. If the spin absorption by the static layer is anisotropic, the spin-pumping contribution is manifested as an angle-dependent modulation of the FMR linewidth. Indeed, anisotropic absorption of pure spin currents was reported for F/N/F$_\mathrm{st}$~\cite{Heinrich2003,Baker2016} but could not be attributed to noncollinearity between the spin-current polarization and magnetization. In Ref.~\onlinecite{Heinrich2003}, it was attributed to the angle dependence of dynamic exchange, whereas in Ref.~\onlinecite{Baker2016}, it was explained in terms of an angular dependence of the total Gilbert damping in F$_\mathrm{st}$. To the best of our knowledge, the anisotropy of spin relaxation studied by controllably varying the angle between the spin-current polarization (set by F) and the magnetic axis of F$_\mathrm{st}$ or AF has not been demonstrated in magnetic multilayers. The main difficulty in achieving a reliable control of the noncollinear alignment in such spin valve structures lies in the presence of a kOe-range external magnetic field required in a typical FMR experiment, which often fully aligns the studied multilayer magnetically.

In this Letter, we demonstrate controllable magnetic damping in a F layer via $\phi$-dependent interaction of the emitted spin pumping current with the N\'eel vector of an AF layer in a F/N/AF/F$_\mathrm{p}$ type spin-valve, where $\phi$ is the angle between the equilibrium magnetization in F and the AF N\'eel vector. Carrying out detailed $\phi$-dependent, variable-temperature measurements of the FMR-driven, spin-pumping-mediated magnetization damping, we observe a pronounced maximum when the magnetization of F is orthogonal to the AF N\'eel vector. Our results are well described by a theoretical model analogous to that for ferromagnetic spin valves~\cite{Akash2018} and indicate the dominance of \emph{longitudinal} over spin-mixing conductance in spin relaxation by the AF layer. This consistence with the recent prediction~\cite{Akash2018} highlights the importance of longitudinal spin transport in magnetic multilayers and establishes an important pathway towards achieving in-situ damping tunability. 

\begin{figure}
\includegraphics[width=9 cm]{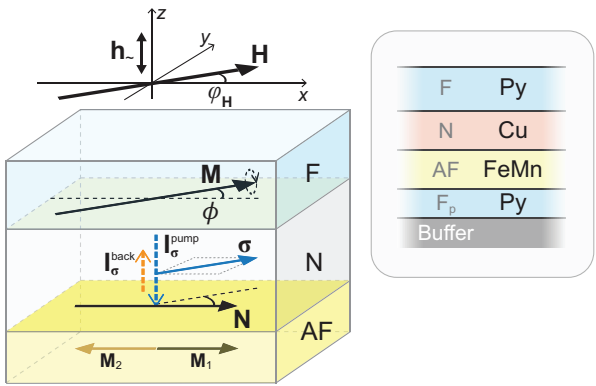}
\caption{Schematic of the studied multilayer and the FMR measurement configuration when external magnetic field $\mathbf{H}$ is applied in the film plane $xy$. The equilibrium axis of resonating magnetization $\mathbf{M}$ in F follows the external field $\mathbf{H}$, whereas the AF vector $\mathbf{N}$ is fixed in the $xy$ plane  (using e.g. a ferromagnetic seed layer, F$_\mathrm{p}$; inset). Vector $\mathbf{N}$ forms angle $\phi$ with polarization $\bm{\sigma}$ of spin-pumped current $\mathbf{I}_\mathrm{s}^\mathrm{pump}$.}
\label{fig_1}
\end{figure}

\paragraph{Spin pumping experiment.} Spin pumping is an emission of spin angular momentum ($\mathbf{I}_{\bm{\sigma}}^\mathrm{pump}$) by a resonantly precessing ferromagnet (F) into an adjacent nonmagnetic spacer N~\cite{Tserkovnyak2002}. In this sense, spin pumping is a reciprocal effect to a spin-transfer torque~\cite{Tserkovnyak2005}. $\mathbf{I}_{\bm{\sigma}}^\mathrm{pump}$ carries spin away from F, which increases the magnetization damping in F, usually detected as a broadening of the F-layer's FMR linewidth~\cite{Mizukami2001,Foros2005}. Considering the F/N/AF trilayer used in this work, and taking spin relaxation in N to be negligible, a fraction of the spin-pumping current is reflected at the N/AF interface and returns back to F. The spin-pumping contribution to the FMR linewidth is proportional to the difference between $\mathbf{I}_{\bm{\sigma}}^\mathrm{pump}$ and the back-flow spin current $\mathbf{I}_{\bm{\sigma}}^\mathrm{back}$ (Fig.~\ref{fig_1})~\cite{Tserkovnyak2005}. A change in the relative orientation between the spin-current polarization $\bm{\sigma}$ and the AF's Neel vector should affect $\mathbf{I}_{\bm{\sigma}}^\mathrm{back}$ thereby modulating the FMR-linewidth of F.

The F/N/AF trilayer under ``in-plane'' FMR, illustrated in Fig.~\ref{fig_1}, has the equilibrium orientation of the resonating magnetization in magnetically soft F ($\mathbf{M}$) aligned with the external magnetic field ($\simeq$1 kOe) applied at angle $\varphi_\mathbf{H}$. At the same time, the N\'eel vector $\mathbf{N}$ of AF is essentially insensitive to this relatively weak field and remains directionally fixed, provided it is suitably set in fabrication (as detailed below). This spin valve structure allows to control the angle $\phi$ between the F layer magnetization and the AF layer N\'eel vector enabling us to extract the $\phi$-dependence of the spin-pumping contribution to the FMR linewidth.

A macroscopic magnetic anisotropy in AF -- the key AF property for this study -- can be induced by deposition in a static magnetic field and/or post-fabrication magnetic annealing~\cite{Nogues1999}. The other effective approach is to deposit a thin AF layer onto a saturated ferromagnetic seed layer (F$_\mathrm{p}$), which induces a strong exchange bias in the AF/F$_\mathrm{p}$ bilayer. The latter results in a pronounced unidirectional magnetic anisotropy in F$_\mathrm{p}$ as well as a magnetic axis in AF. We have fabricated a series of multilayers, where antiferromagnetic FeMn is grown on either a ferromagnetic (Py) or nonmagnetic  (Ta) seed layer. Here, FeMn and Py denote Fe$_{50}$Mn$_{50}$ and Fe$_{20}$Ni$_{80}$ (permalloy) alloys. The thickness of the FeMn layer of 7~nm was found to be optimal as regards to a strong directional exchange bias throughout the FeMn/Py bilayer with a high blocking temperature ($T_\mathrm{b} \approx 420$~K)~\cite{Jungblut1994,Nogues1999}. The opposite surface of the FeMn layer, acting as the spin-current reflector or sink, is interfaced with the free, soft Py layer via a nonmagnetic Cu spacer. The thickness of the Cu layer (6~nm) was chosen much smaller than the spin diffusion length in Cu ($\lambda_\mathrm{s} > $ 100~nm at room temperature~\cite{Bass2007}) to ensure negligible spin dissipation in N.

\begin{figure}
\includegraphics[width=9 cm]{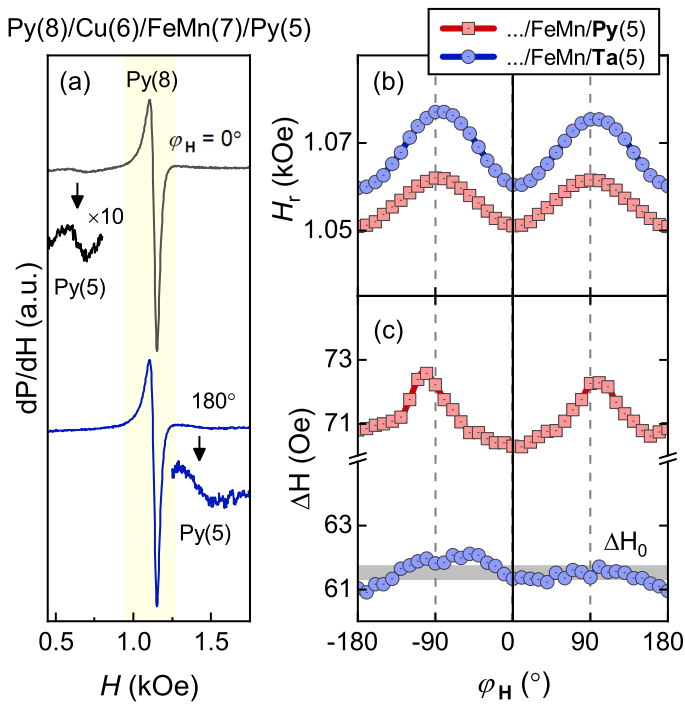}
\caption{(a) In-plane FMR spectra measured along ($0\degree$) and opposite ($180\degree$) to the exchange-pinning direction set by the seed Py(5) layer. (b), (c) Angular dependence of the resonance field ($H_\mathrm{r}$) and the linewidth ($\Delta H$) for the free Py(8) layer for the structures where the antiferromagnetic FeMn layer was grown on nonmagnetic Ta(5) or ferromagnetic Py(5). The data were measured at 280~K.}
\label{fig_2}
\end{figure}

FMR measurements  were carried out at a constant frequency of 9.88~GHz while sweeping an external magnetic field $\mathbf{H}$ applied in the film plane. The obtained spectra exhibit a strong resonance line from the free Py layer, the position of which (the resonance field) reveals a very weak uniaxial magnetic anisotropy in the film plane [Figs.~\ref{fig_2}(a) and (b)]. The FMR spectra for the structures with the FeMn/Py bilayer exhibit an additional weakly-intensive resonance line, which we attribute to the seed Py layer and use as an independent probe of the magnetic directionality of AF. This line exhibits a very pronounced unidirectional anisotropy, indicating a strong exchange bias in the FeMn/Py bilayer~\citep{Nogues1999}. 

\paragraph{Angle-dependent FMR linewidth.} The FMR linewidth of the free Py layer $\Delta H$ as well as the resonance field $H_\mathrm{r}$ were obtained by fitting the spectra with a Dysonian~\cite{Dyson1955}. $\Delta H$ versus $\varphi_\mathbf{H}$ for the samples with the magnetic and nonmagnetic seed layers differ significantly in the magnitude of the variation as well as its angular profile [Fig.~\ref{fig_2}(c)]. On the other hand, the respective resonance fields $H_\mathrm{r}(\varphi_\mathbf{H})$ show the same behavior [Fig.~\ref{fig_2}(b)], indicating that the saturation magnetization and the magnetic anisotropy of the free layer are largely unaffected by the seed layer. The observed difference in the magnetic damping ($\Delta H$) for the two structures must therefore be attributed to a difference in the magnetic state of their respective FeMn layers. The essentially isotropic behavior of $\Delta H$ for the case of FeMn/Ta indicates that the FeMn layer exhibits no macroscopic magnetic axis (in fact, by design) as antiferromagnetic domains formed on non-magnetic Ta are likely to orient randomly. In contrast, the pronounced anisotropy of $\Delta H$ for the structure with the FeMn/Py bottom magnetic layer is evidence for a well-defined macroscopic magnetic anisotropy in the FeMn layer, the conclusion additionally and independently supported by the observed pronounced unidirectional anisotropy in the seed-Py layer~\cite{SupplMater}. In what follows, we focus on this key result of anisotropic spin relaxation in the FeMn/Py-based structure.

The multilayers with FeMn/Py exhibit a thermally-induced transformation of the $\Delta H$-vs-$\varphi_\mathbf{H}$ profile. The maximum in $\Delta H(\varphi_\mathbf{H})$ observed at room temperature at $\varphi_\mathbf{H} \approx \pm 90\degree$ shifts to larger angles with decreasing temperature [Fig.~\ref{fig_3}(a)]. At the same time, there are no temperature-induced changes in the angle profiles of the resonance field, except the offset in the magnitude of $H_\mathrm{r}$ due to the temperature variation in the saturation magnetization of the Py layer~\cite{SupplMater}. This implies that the observed changes in $\Delta H$-vs-$\varphi_\mathbf{H}$ with temperature are caused by factors external to the free layer -- in our case the spin-pumping-mediated effect of the AF on $\Delta H$ -- rather than by any changes in the intrinsic magnetic properties of the free layer. Importantly, the changes in the angular profiles of Fig.~\ref{fig_3}(a) are associated with a temperature dependence of the AF layer pinning by the seed underlayer. This is strongly supported by the corresponding $\Delta H$-vs-$\varphi_\mathbf{H}$ for the FeMn/Ta-based structures (without macroscopic magnetic axis), which remain largely isotropic at all temperatures~\cite{SupplMater}.

\paragraph{Spin-pumping contribution.} We explain the observed angle-dependence of the FMR linewidth as due to the spin-pumping contribution to the magnetization dynamics of the free layer. We extend the phenomenology of the spin-pumping effect~\cite{Tserkovnyak2002,Tserkovnyak2005} to a noncollinear ferromagnetic-nonmagnetic-antiferromagnetic (F/N/AF) trilayer system. The magnetization dynamics of F is described {in terms of the spin conductivities of the F/N and N/AF interfaces for an arbitrary mutual orientation of the respective magnetic order parameters -- magnetization $\mathbf{M}$ and N\'eel vector $\mathbf{N}$.

The anisotropic $\Delta H$-vs-$\varphi_\mathbf{H}$ profile, shown in Fig.~\ref{fig_2}(c), exhibits maxima close to $\varphi_\mathbf{H} = \pm 90\degree$, which can be qualitatively explained as follows. The precessing F pumps spins into the adjacent spacer layer, a fraction of which is subsequently absorbed by AF. The absorbed spin current manifests as additional damping in F. However, only the component of spin current that is orthogonal to the equilibrium F magnetization determines the damping in it. This component is best absorbed when the F magnetization and the AF N\'eel vector are mutually orthogonal, provided that AF absorbs and dissipates the longitudinal component stronger than the transverse component of the spin current.

In the general case, the spin-pumping contribution to the integral FMR linewidth can be quantitatively expressed in terms of the spin conductances of the F/N and N/AF interfaces~\cite{Akash2018}. The other contributions, such as the intrinsic Gilbert damping and inhomogeneous terms, have a much weaker dependence on the in-plane angle and add up in the total $\Delta H(\varphi_\mathbf{H})$ to a constant $\Delta H_\mathrm{int} \approx \mathrm{const}$~\cite{Kravets2016}. The case of a noncollinear mutual alignment of $\mathbf{M}$ and $\mathbf{N}$ can be treated in a manner analogous to Ref.~\onlinecite{Akash2018}. The full FMR linewidth becomes

\begin{equation}
\Delta \widetilde{H} = \Delta \widetilde{H}_0  + 0.5\frac{\tilde{g}_r\tilde{g}_l}{(\tilde{g}_r \tilde{g}_l + \tilde{g}_r) + (\tilde{g}_l - \tilde{g}_r) \cos^2\phi},
\label{LW1}
\end{equation}

\noindent where $\Delta \widetilde{H} = \Delta H \cdot (4 \pi M V)/(g_r^\star \cdot \hbar \omega)$, $\omega$ -- the frequency of the applied microwave field, $V$ -- the film volume; $\tilde{g}_{l,r} = g_{l,r}/g_r^\star$, where the longitudinal spin conductance ($g_l$) and the real part of the spin-mixing conductance ($g_r$) characterise the N/AF subsystem, whereas $g_r^\star$ relates to the F/N interface. The second term in (\ref{LW1}) is the angle-dependent spin-pumping contribution, a function of angle $\phi$ between $\mathbf{M}$ and $\mathbf{N}$. $\Delta \widetilde{H}_0$, in turn, consists of $\Delta \widetilde{H}_\mathrm{int}$ and the \emph{angle-independent} spin-pumping contribution: $\Delta \widetilde{H}_0 = \Delta \widetilde{H}_\mathrm{int} + 0.5 \tilde{g}_r/(1+\tilde{g}_r)$.

\begin{figure}
\includegraphics[width=9 cm]{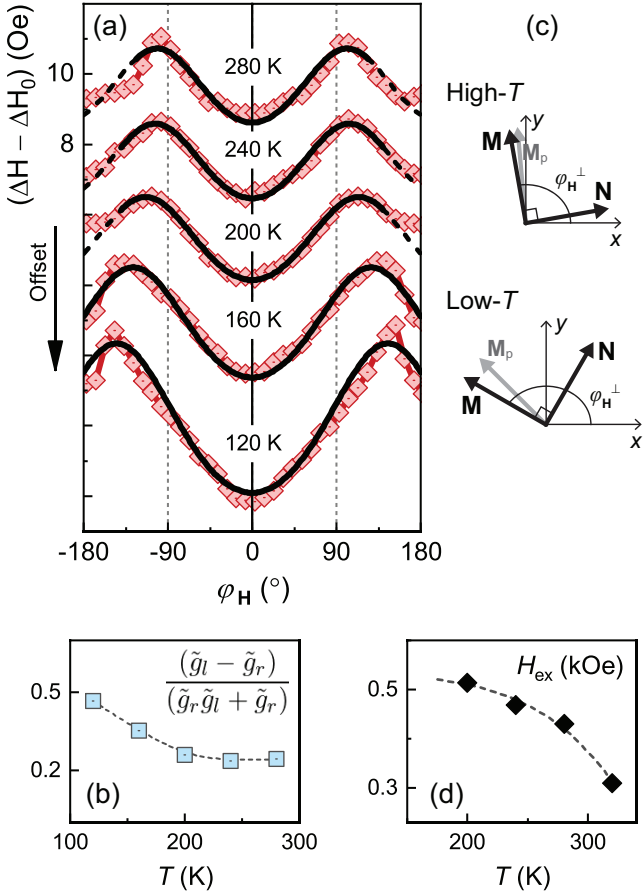}
\caption{(a) Angle-dependent spin-pumping contribution $(\Delta H - \Delta H_0)$ versus $\varphi_\mathbf{H}$, obtained from the measured FMR linewidth by subtracting the predominantly isotropic linewidth for the control sample (FeMn/Ta based) at each given temperature. Solid lines are the fits to theory [Eq.~(\ref{LW1})]. (b) Extracted from the fitting in (a), the temperature dependence of the spin-relaxation asymmetry parameter. (c) Illustration of the rotating torque on the AF with N\'eel vector $\mathbf{N}$ by the exchange-coupled Py(5) layer of magnetization $\mathbf{M}_\mathrm{p}$, for two characteristic temperatures.  (d) FeMn/Py exchange-pinning field $H_\mathrm{ex}$ vs temperature, extracted from $H_\mathrm{r} (T)$ for Py(5) [Fig.~\ref{fig_2}(a)].}
\label{fig_3}
\end{figure}

Equation~(\ref{LW1}) can be used for fitting the experimental $\Delta H$-vs-$\varphi_\mathbf{H}$ data. Subtracting the \emph{angle-independent} background $\Delta H_0$ from the total $\Delta H$ allows one to fit the data using only the second, angle-dependent term in~(\ref{LW1}). In this respect, and based on the detailed discussion above [Figure~\ref{fig_2}(c) and related text], the most appropriate is to take as $\Delta H_0$ the linewidth for the FeMn/Ta-based structure, in fact, specifically designed for this calibration. Figure~\ref{fig_3}(a) shows the result, which agrees well with the experiment. 

\paragraph{Effect of temperature.} With changing temperature, the anisotropic spin relaxation undergoes a transformation of its angular form [Fig.~\ref{fig_3}(a)], which can be explained by the temperature-dependent properties of the exchange-pinned AF/F$_\mathrm{p}$ bilayer. The stronger interface exchange-pinning at lower temperatures results in a stronger torque on the AF, such that $\mathbf{M}$ and $\mathbf{N}$ become orthogonal at different angles of the applied in-plane field (of fixed magnitude $H = H_\mathrm{r}^\mathrm{F} =$ 1.03--1.08~kOe) for different temperatures, as detailed below. 

With decreasing temperature, the maximum in $\Delta H$ shifts from $\varphi_\mathbf{H} \approx \pm 90\degree$ to larger angles [Fig.~\ref{fig_3}(a)], which can be explained by a deviation of vector $\mathbf{N}$ (tilt angle $\varphi_\mathbf{N}$) from its easy (exchange-pinning) direction, so the 90$\degree$-rotation of $\mathbf{M}$ with respect to $\mathbf{N}$ occurs at a larger $\varphi_\mathbf{H}$ [field angle measured from the exchange-pinning direction, as illustrated in Figs.~\ref{fig_3}(c)]. This tilt of $\mathbf{N}$, increasing at low temperatures for a given field-torque (acting via F$_\mathrm{p}$), is due to the well-known strengthening of the exchange coupling between AF and F$_\mathrm{p}$ (increasing from about AF's $T_\mathrm{N}$ toward low temperature). The magnetization $\mathbf{M}_\mathrm{p}$ of F$_\mathrm{p}$ follows the direction of the applied magnetic field $\mathbf{H}$ (the applied ~1 kOe exceeds the exchange-pinning field at all temperatures) and, via the exchange at the interface, torques $\mathbf{N}$ off the initial equilibrium orientation ($\varphi_\mathbf{N} = 0$). This tilting is quantitatively described by a competition between the exchange bias in the AF/F$_\mathrm{p}$ bilayer ($J_\mathrm{ex}$) and the Zeeman energy of F$_\mathrm{p}$ ($J_\mathrm{Z}$), as detailed in~\cite{SupplMater}. With changing temperature, $J_\mathrm{Z}$ varies slowly since the Curie point of F$_\mathrm{p}$ is much higher than the experimental temperature range, whereas $J_\mathrm{ex}$ has a pronounced temperature dependence shown by the extracted exchange field $H_\mathrm{ex} (\sim J_\mathrm{ex})$ vs $T$ in Fig.~\ref{fig_3}(d): $H_\mathrm{ex} \approx 0.5$~kOe vanishes toward the Neel point of the AF (more precisely the blocking point, $T_\mathrm{b} \approx T_\mathrm{N}$~\cite{Nogues1999}). 

Equation (\ref{LW1}) fits the measured data very well for all temperatures [solid lines in Fig.~3(a)], when modified according to the discussion above, such that angle $\phi$ between $\mathbf{M}$ and $\mathbf{N}$ is scaled by a temperature-dependent parameter reflecting the tilt of the AF, $\phi = (\varphi_\mathbf{H} - \varphi_\mathbf{N}) = a \varphi_\mathbf{H}$ ($a \leq 1$). Our simulations show~\cite{SupplMater}, that the AF tilt angle $\varphi_\mathbf{N}$ is linear in $\varphi_\mathbf{H}$ in the angle interval slightly wider than that between the two maxima in $\Delta H(\varphi_\mathbf{H})$. The parts of the calculated curves outside this fitting interval are dashed in Fig.~\ref{fig_3}(a). The final result of the analysis is the parameter representing the anisotropic magnetization damping, which is proportional to the difference $(g_l - g_r)$ [Fig.~\ref{fig_3}(b)] and shows how the observed angle-dependent FMR linewidth directly stems from the spin-conductance asymmetry of the N/AF interface.

\paragraph{Discussion and Conclusion.} The result of the above analysis is a finite and positive $(g_l - g_r)$ [Fig.~\ref{fig_3}(b)], which means that the spin reflection at the N/AF interface is larger when the spin-current polarization $\bm{\sigma}$ is collinear with the AF's magnetic axis ($\mathbf{x}$) and smaller for orthogonal $\bm{\sigma}$ and $\mathbf{x}$. In arriving at this conclusion, we have assumed that the exchange biasing tends to align the N\'eel vector of the AF with the seed layer's magnetization. This assumption is supported by the widely accepted \emph{uncompensated-spin} model of exchange bias~\cite{Malozemoff1987,Mauri1987} describing the effect in similar metallic AF/F$_\mathrm{p}$ bilayers~\cite{Antel1999}. Therefore, our experiment demonstrates that $g_l$ exceeds $g_r$ in the considered metallic AFs, which is consistent with the strong spin relaxation observed in such AFs~\cite{Merodio2014,Frangou2016}, and is in contrast to the typical assumption, $g_l \ll g_r$, in literature~\cite{Tserkovnyak2003,Tserkovnyak2005,Taniguchi2007}. 

Furthermore, the shown sensitivity of the magnetization damping in the resonating F layer to the presence of an induced magnetic axis in the static AF layer is an example of how spin-pumping can be used for probing the changes in the spin configuration of AFs subjected to external stimuli (thermal and/or magnetic). Finally, the reported angular modulation of the FMR linewidth is experimental demonstration of an in-situ, spin-pumping-mediated control of magnetization damping in magnetic multilayers predicted recently~\cite{Akash2018}.

\begin{acknowledgments}
Support from the Swedish Research Council (VR Grant No.~ 2018-03526), the Swedish Stiftelse Olle Engkvist Byggm\"astare, and the Research Council of Norway through its Centers of Excellence funding scheme, Project No. 262633, “QuSpin”, are gratefully acknowledged.
\end{acknowledgments}

\bibliography{References}

\newpage
\onecolumngrid
\appendix

\section{SUPPLEMENTAL MATERIAL}

\subsection{NOTE 1. Resonance field of the free Py(8) layer as a function of temperature}

\begin{figure}[h]
\includegraphics[width=9 cm]{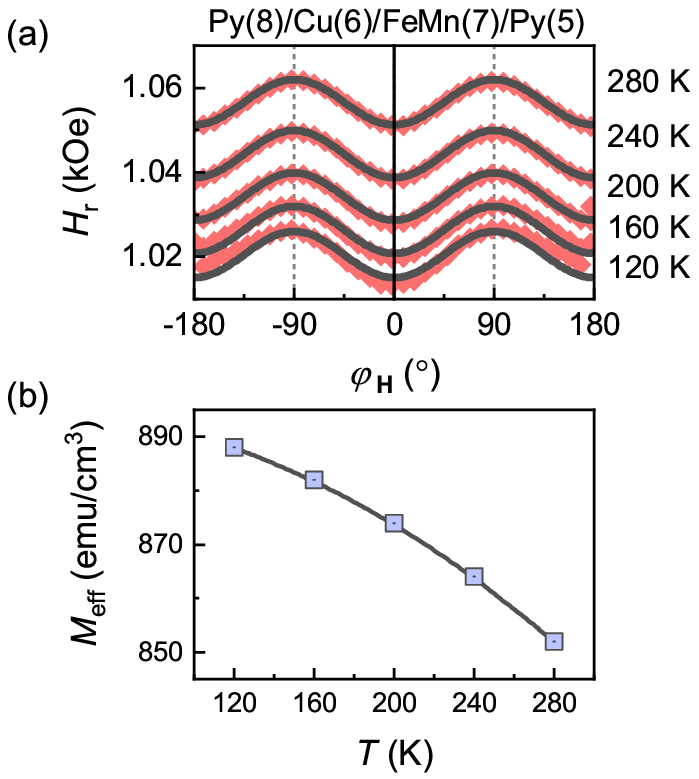}\\ 
\textbf{Figure A1.} (a) Resonance field of the free Py(8) layer as a function of temperature. (a) In-plane angular profile of the resonance field ($H_\mathrm{r}$) shows the same weak uniaxial magnetic anisotropy at all temperatures. The slight increase in the magnitude of $H_\mathrm{r}$ is due to the thermal demagnetization (decrease in the effective magnetization, $M_\mathrm{eff}$) of the free Py layer, shown in panel (b). The values of $M_\mathrm{eff}$ at the given measurement temperatures were obtained from fitting the $H_\mathrm{r}$-vs-$\varphi_\mathbf{H}$ profiles [solid lines in panel (a)] using the standard FMR phenomenology [J. Smit and H. G. Beljers, Phillips Res. Rep. 10, 113 (1955)].
\label{fig_A1}
\end{figure}

\newpage
\subsection{NOTE 2. Resonance field and FMR linewidth versus temperature}

\begin{figure}[h]
\includegraphics[width=10 cm]{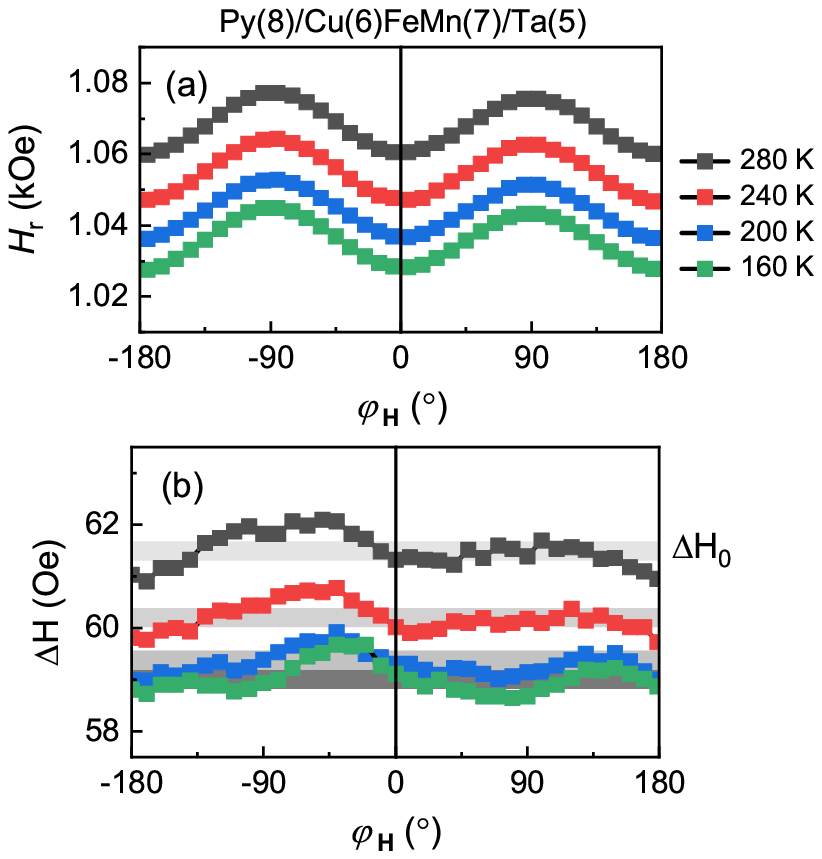}\\ 
\textbf{Figure A2.} Resonance field and FMR linewidth versus temperature for the structure with a nonmagnetic seed layer (with FeMn/Ta bilayer). (a) In-plane angular profiles of the resonance field of the free Py layer show a similar behavior to that for the structure with a magnetic seed layer [with FeMn/Py bilayer; see Fig.~A1(a)]. (b) FMR linewidth ($\Delta H$) shows a much less pronounced angular dependence at all temperatures, which is in contrast to the significant temperature induced changes in $\Delta H$ for the structure with FeMn/Py [Figs.~\ref{fig_2}(c),~\ref{fig_3}(a)].
\label{fig_A2}
\end{figure}

\newpage
\subsection{NOTE 3. Rotation of the N\'eel vector of the AF}

\begin{figure}[h]
\includegraphics[width=10 cm]{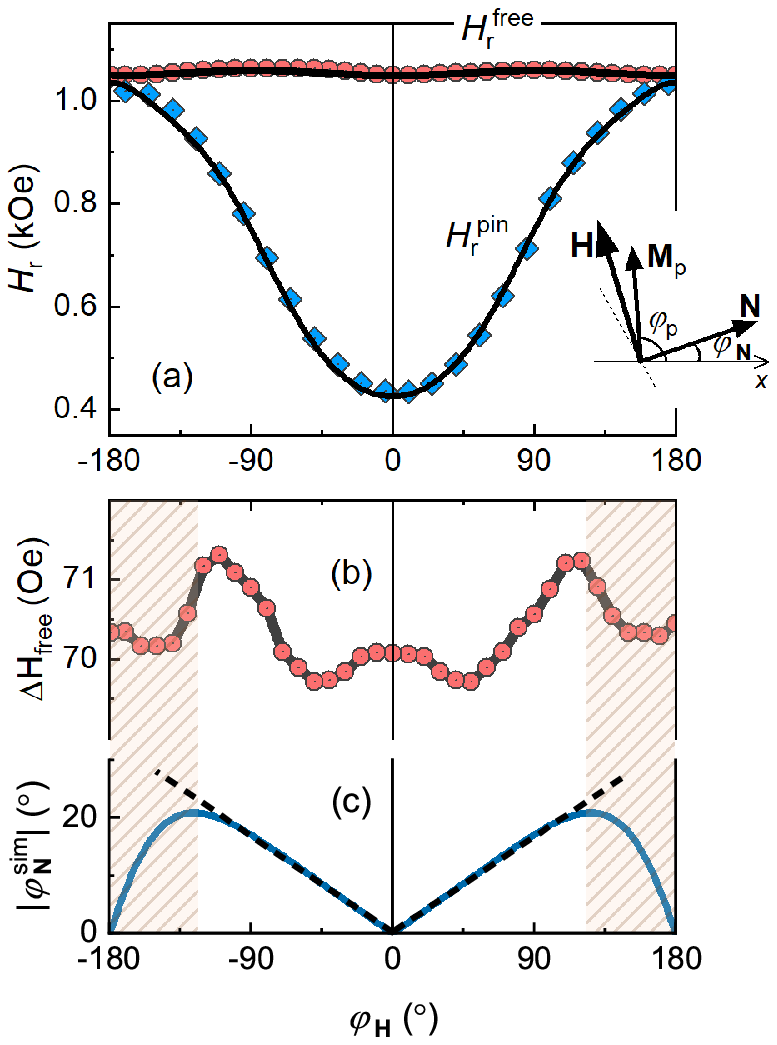}\\ 
\textbf{Figure A3.} Rotation of the N\'eel vector of the AF extracted from the angular dependences of the resonance field using the model fit detailed in Note~1. (a) In-plane angular dependences of the resonance field for the free Py layer ($H_\mathrm{r}^\mathrm{free}$) and the pinned Py layer ($H_\mathrm{r}^\mathrm{pin}$) of the Py(8)/Cu(5)/FeMn(5)/Py(5) structure, measured at 280~K. The lines are the fits performed using the model of Note~4. (b) Corresponding angular dependences of the FMR linewidth of the free Py layer ($\Delta H_\mathrm{free}$) and (c) simulated misalignment of the N\'eel vector ($\varphi_\mathbf{N}^\mathrm{sim}$) relative to the exchange-pinning direction ($\varphi_\mathbf{H} = 0\degree$) set at fabrication in the FeMn/Py bilayer. The dependence in (c) was calculated using the parameters obtained from the model fit shown in (a). The misalignment $\varphi_\mathbf{N}^\mathrm{sim}$ is linear in $\varphi_\mathbf{H}$ in the angle interval between the two maxima in $\Delta H_\mathrm{free}$-vs-$\varphi_\mathbf{H}$, shown in (b).
\label{fig_A3}
\end{figure}

\newpage
\subsection{NOTE 4. Gibbs free energy for F/AF bilayers}

To model the deviation of the N\'eel vector of the F/AF bilayer from its easy (exchange-pinning) direction, we split the energy term commonly used for describing exchange bias into two terms, corresponding to the interlayer exchange coupling at the F/AF interface and the magnetic anisotropy in AF. The Gibbs free energy per unit volume becomes

\noindent where $J_\mathrm{F}$ and $J_\mathrm{AF}$ are the constants of the uniaxial magnetic anisotropy in the pinned ferromagnetic Py and the antiferromagnetic FeMn layers, respectively; $J_\mathrm{ex}$ is the constant of exchange coupling at the Py/FeMn interface; $Z$ and $G_\mathrm{dem}$ are the Zeeman and demagnetization energies; $\mathbf{m}_\mathrm{p}$ is the unit vector along magnetization $\mathbf{M}_\mathrm{p}$ of the pinned Py.

\begin{equation}\tag{A1}
\begin{aligned}
G_\mathrm{p} = Z~+ &~G_\mathrm{F} + G_\mathrm{AF} + G_\mathrm{ex} + G_\mathrm{dem} = \\
& = -\mathbf{M}_\mathrm{p} \mathbf{H} - J_\mathrm{F}(\mathbf{x}\cdot\mathbf{m}_\mathrm{p})^2 - J_\mathrm{AF}(\mathbf{x}\cdot\mathbf{N})^2 - J_\mathrm{ex}(\mathbf{m}_\mathrm{p}\cdot\mathbf{N}) + 2\pi M_\mathrm{p}^2 (\mathbf{z}\cdot\mathbf{m}_\mathrm{p})^2.
\end{aligned}
\end{equation}

When external magnetic field $\mathbf{H}$ is oriented in the film plane, terms $G_\mathrm{F}$, $G_\mathrm{AF}$, and $G_\mathrm{ex}$ can be transformed as follows: $G_\mathrm{F} = -0.5 M_\mathrm{p}H_\mathrm{ua}\cos 2 \varphi_\mathrm{p}$, $G_\mathrm{AF} = -J_\mathrm{AF} \cos \varphi_\mathbf{N}$ and $G_\mathrm{ex} = -M_\mathrm{p}H_\mathrm{ex}\cos (\varphi_\mathrm{p}-\varphi_\mathbf{N})$, where angles $\varphi_\mathrm{p}$ and $\varphi_\mathbf{N}$ are defined in the inset to Fig.~A3(a). Minimization of (A1) gives

\begin{equation}\tag{A2}
\begin{cases}
\dfrac{J_\mathrm{AF}}{J_\mathrm{ex}} \sin( 2\varphi_\mathbf{N}) = \sin (\varphi_\mathrm{p}-\varphi_\mathbf{N}), \\[3mm]
\dfrac{Z}{J_\mathrm{ex}} \sin (\varphi_\mathrm{p}-\varphi_\mathbf{H}) = -\sin (\varphi_\mathrm{p}-\varphi_\mathbf{N}),
\end{cases}
\end{equation}

\noindent where $Z = H M_\mathrm{p}$.

Equations (A1) and (A2) can be used for fitting the angular dependence of the FMR spectra of the pinned Py layer in the general case of non-rigid $\mathbf{N}$. Curve $H_\mathrm{r}^\mathrm{pin}(\varphi_\mathbf{H})$ in Fig.~A3(a) is an example of such fitting, showing a good agreement between the fit and the experimental data.

\end{document}